\documentclass{ws-procs9x6}

\begin{document}

\title{Electroweak physics and physics beyond the Standard Model}

\author{L.~Bellagamba}
\address{INFN Bologna, \\
       Via Irnerio 46, I-40126 Bologna, Italy, \\
       E-mail: Lorenzo.Bellagamba@bo.infn.it}
\author{E.~Sauvan}
\address{CPPM, IN2P3-CNRS et Universit\'e de la M\'editerran\'ee, \\
       163 av.\ de Luminy F-13288 Marseille, France, \\
       E-mail: sauvan@cppm.in2p3.fr}
\author{H.~Spiesberger}
\address{Johannes-Gutenberg-Universit\"at Mainz, \\ 
       Staudinger Weg 7, D-55099 Mainz, \\
       E-mail: hspiesb@thep.physik.uni-mainz.de}

\maketitle

\abstracts{We summarize the recent results on electroweak physics and
  physics beyond the Standard Model that have been presented at the
  XIV International Workshop on Deep Inelastic Scattering 2006.}


\section{Top Mass and Electroweak Precision Measurements}

\begin{figure}[tb]
\centerline{\epsfxsize=6cm\epsfbox{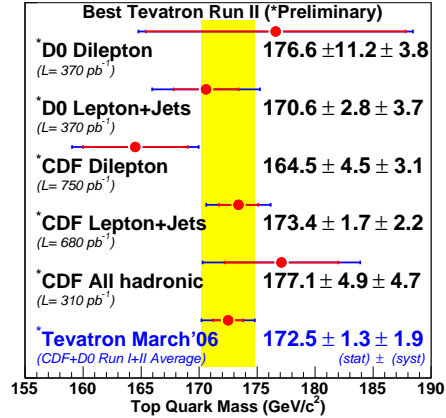}}
\caption{Compilation of top mass measurements at the
  Tevatron. \label{topmass}}
\end{figure}

A precise measurement of the top mass is a central goal of the CDF and
D0 experiments which, by 2006, collected an integrated luminosity of
$\sim 1$~fb$^{-1}$ during the Tevatron run II operations.  The top mass
enters as an important parameter in the calculation of electroweak (EW)
observables. In global fits of the EW observables to Standard Model (SM)
parameters, the top mass therefore provides an indirect constraint on
the Higgs mass and therefore plays a crucial role in the understanding
of the SM Higgs sector. It is hence highly desirable to improve the
precision of the top mass measurement in order to reach a precision
comparable to that of the other relevant EW parameters, typically of the
order $0.1\%$ - $1\%$. Figure~\ref{topmass} which shows a compilation of
measurements from both D0 and CDF, witnesses the great progress obtained
in the last few years due to the increased statistics, the upgrade of
the detectors for run II and the use of new analysis techniques which
allow to better control the main systematic uncertainties.

The systematic error is dominated by the uncertainty on the jet energy
scale (JES).  Important progress in reducing the JES uncertainty has
been obtained using the template and the matrix element
methods\cite{talk_topmass}.  The most precise measurement comes from the
lepton+jet channel which gives the best compromise between statistics
and background.  The di-lepton channel gives important contribution
since it is less affected by the JES while the all-hadronic channel
suffers from a large background and its contribution to the top mass
measurement is, at the moment, negligible.  The combination of the
different measurements gives $M_{\rm top} = 172.5 \pm 1.3\mbox{(stat)}
\pm 1.9 \mbox{(sys)}$~GeV.  In the coming years of data taking with
$4$-$8$~fb$^{-1}$ of integrated luminosity foreseen up to the end of
operations at the Tevatron, the di-lepton channel is expected to become
dominated by systematic errors and also the all-hadronic channel will
contribute significantly, allowing to reach a top mass precision of
$\sim 1.5$~GeV.

The Tevatron measurements will provide the most precise top mass value
also during the first years of data taking at the LHC.  Top production
at the LHC experiments will play an important role in the detector
commissioning, since it produces a rich topology with $b$-jets, missing
$P_T$, high-$P_T$ jets and high-$P_T$ leptons. This involves several
detection aspects allowing a global monitoring of the detector
performances\cite{talk_toplhc}. A precision of the order of or better
than the one at the Tevatron could be reached with $10$~fb$^{-1}$ of
integrated luminosity.

\begin{figure}[tb]
\vspace*{-0.8cm}
\centerline{\epsfxsize=6cm\epsfbox{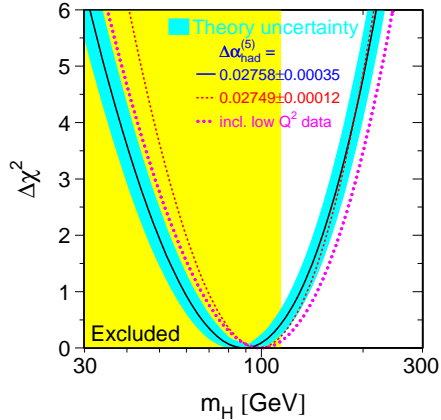}}
\vspace*{-0.7cm}
\caption{Higgs mass dependence of the $\chi^2$ of the electroweak
  parameter fit. \label{higgs}}
\end{figure}

The new Tevatron measurement of the top mass can be used, together with
other precise electroweak measurements, for a global fit which allows to
extract the SM parameters and test the compatibility of the theory with
the present experimental picture. Since many of the EW observables are
sensitive to the top mass via radiative corrections, the Higgs mass
resulting from these global fits is affected by the top mass.
Figure~\ref{higgs} shows the $\chi^2$ of the electroweak fit as a
function of the Higgs mass.  The experimental results clearly favor a
light Higgs; masses above $175$~GeV ($207$~GeV taking into account the
LEP2 limit) are excluded at $95\%$ CL.

\begin{figure}[tb]
\begin{minipage}[t]{5cm}
\centerline{\epsfxsize=6cm\epsfbox{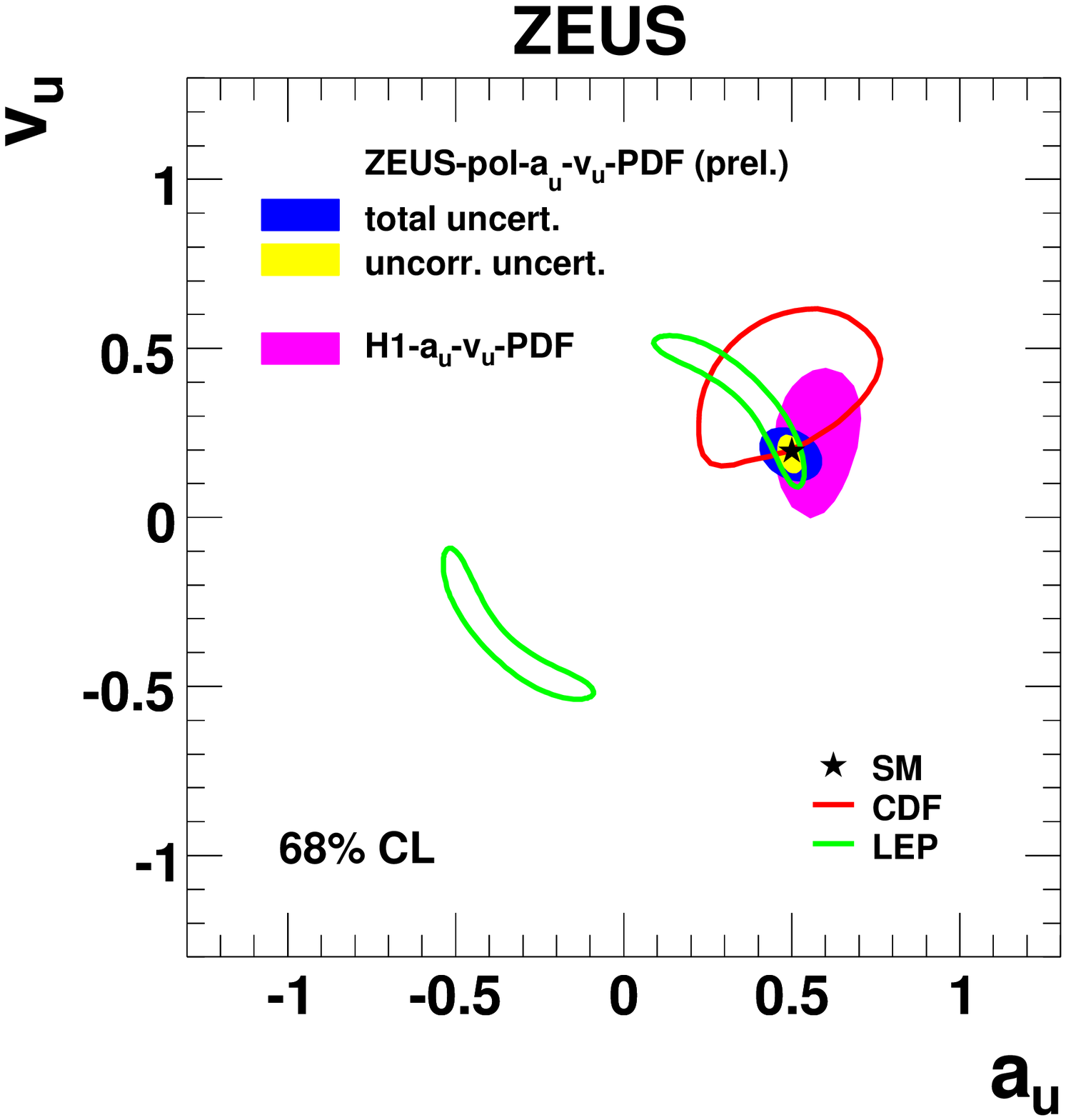}}
\end{minipage} \hfil
\begin{minipage}[t]{5cm}
\centerline{\epsfxsize=6cm\epsfbox{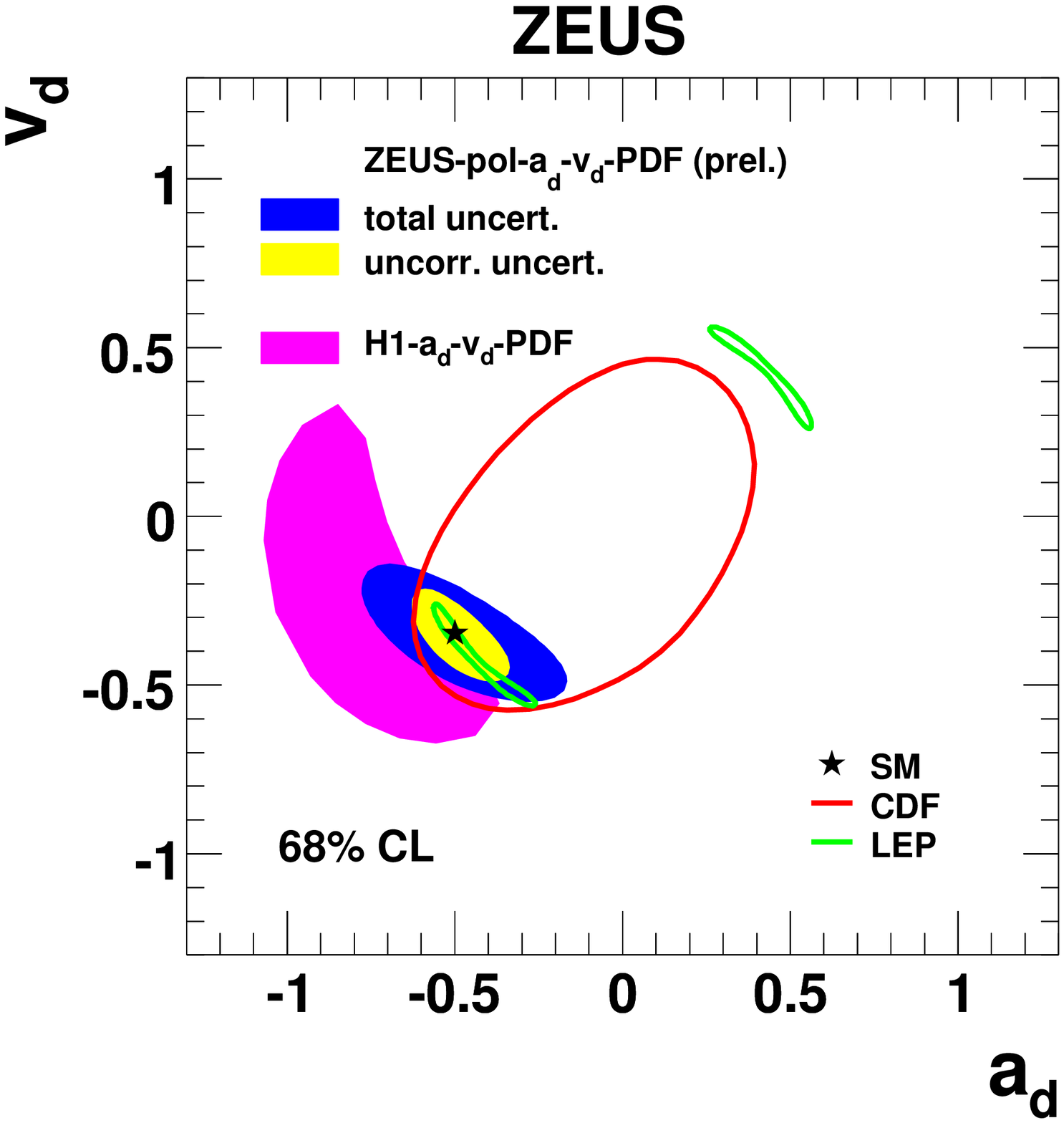}}
\end{minipage} \hfil
\caption{ZEUS constraints for EW NC vector and axial vector couplings of
  the $Z$ boson to quarks. \label{ewzeus}}
\end{figure}

Also the HERA experiments, exploiting polarized electron and positron
beams, have started to contribute to electroweak precision measurements.
In particular a combined electroweak and QCD fit\cite{talk_ewzeus}
allows to extract the EW neutral current (NC) vector and axial vector
couplings of the $Z$ boson to quarks. Figure~\ref{ewzeus} shows the ZEUS
constraints in the plane of the vector ($v_u$, $v_d$) and axial vector
($a_u$, $a_d$) couplings. The results are in agreement with the SM
expectations and competitive with or better than the Tevatron and LEP
constraints.

\section{Higgs Searches}

\begin{figure}[htb]
\centerline{\epsfxsize=8cm\epsfbox{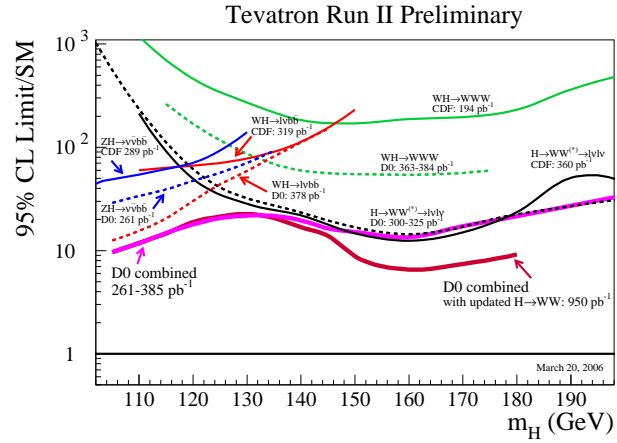}}
\caption{Comparison of the Higgs production limits from CDF and D0 with
  the SM expectations. \label{higgs_tev}}
\end{figure}

In the next few years, with the increasing integrated luminosity
delivered by the Tevatron and the start-up of the LHC, the range of mass
values where the SM Higgs is expected will be accurately scanned.  At
the Tevatron the Higgs\cite{talk_higgstev} can be singly produced via
gluon-gluon ($gg$) fusion or in association with a $W$ or a $Z$ boson in
$q\bar{q}$ annihilation. The $gg$ fusion is typically a factor of 5
larger, but for a light Higgs ($M_{H} \lesssim 135$~GeV), when $H
\rightarrow b\bar{b}$ is the dominant decay, the Higgs signal is
overwhelmed by the QCD background. In this case the associate production
with a weak boson allows to control the background requiring
high-$P_T$ leptons and/or missing $P_T$. If the Higgs is heavier,
the decay into a couple of weak bosons becomes dominant and single Higgs
production via $gg$ fusion is the most sensitive channel. All this is
illustrated in Figure~\ref{higgs_tev} where the present $95\%$~CL limits
from CDF and D0 are compared with the SM expectations. At the moment,
for $M_{\rm H}=115$~GeV, i.e.\ just above the LEP2 exclusion limit, the
sensitivity is $\sim 15$ times above the SM expectation. At least
$2$~fb$^{-1}$ are needed to start excluding part of the mass range and
an integrated luminosity of $8$~fb$^{-1}$ will allow to cover all the
mass range up to $180$~GeV.

\begin{figure}[htb]
\centerline{\epsfxsize=6cm\epsfbox{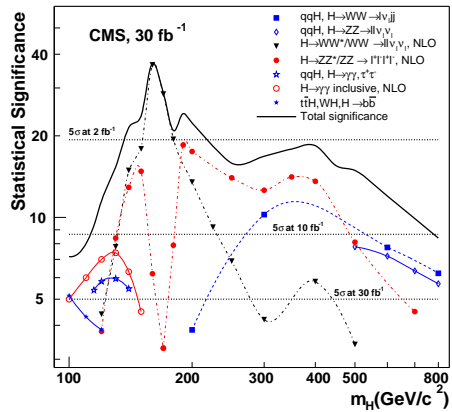}}
\caption{CMS perspectives for a possible Higgs discovery as a function of 
  the mass and for different integrated luminosities. \label{higgs_lhc}}
\end{figure}

Also the LHC experiments, Atlas and CMS, will participate in the
competition for a possible Higgs discovery\cite{talk_higgslhc} in the
first years of data taking.  In any case, at the LHC the experimental
conditions are particularly demanding for a light Higgs just above the
LEP2 limit. In this case the most relevant channels are $H \rightarrow
\gamma \gamma$ and the associated production of the Higgs with a top
pair ($H t\bar{t}$, $H \rightarrow b\bar{b}$). A corresponding analysis
requires a very accurate detector calibration which will be difficult to
reach in the early data taking period. The signal detection will be
significantly less challenging for larger Higgs masses.
Figure~\ref{higgs_lhc} shows the CMS perspectives for a possible
discovery as a function of the Higgs mass and for different integrated
luminosities. This picture shows that one may expect strong competition
between the Tevatron and LHC experiments at the end of 2009.

\section{Top and Weak Boson Production}

\begin{figure}[htb]
\centerline{\epsfxsize=6cm\epsfbox{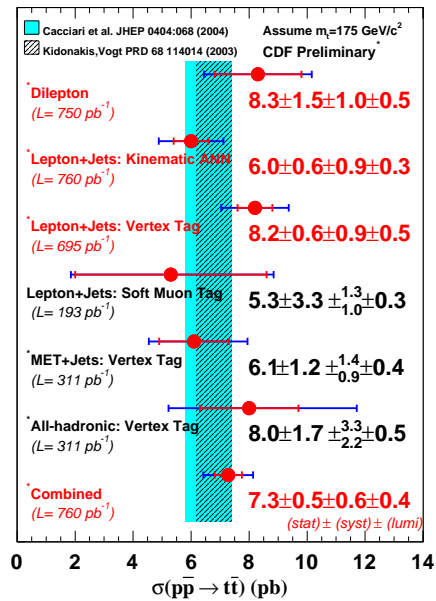}}
\caption{Compilation of measurements of the top production cross
  section at the Tevatron by the CDF Collaboration. Theoretical
  predictions based on NLO calculations are also shown.
  \label{top_cross}}
\end{figure}

Recently, important progress has been made also in the precision of
measurements of the top production cross section\cite{talk_topcross}.
The most precise measurements come from the lepton+jet channel; but, as
illustrated in Figure~\ref{top_cross} showing a compilation of CDF
results, also the other channels give a sizable contribution.  The CDF
combined result is $\sigma_{t} = 7.3 \pm 0.5 \mbox{(stat)} \pm 0.6
\mbox{(sys)} \pm 0.4 \mbox{(lumi)}$~pb, $15\%$ better than the most
precise single measurement. The results are in good agreement with
theoretical calculations at NLO (also shown in the figure).  The D0
collaboration obtained comparable results.

\begin{figure}[tb]
\centerline{\epsfxsize=8cm\epsfbox{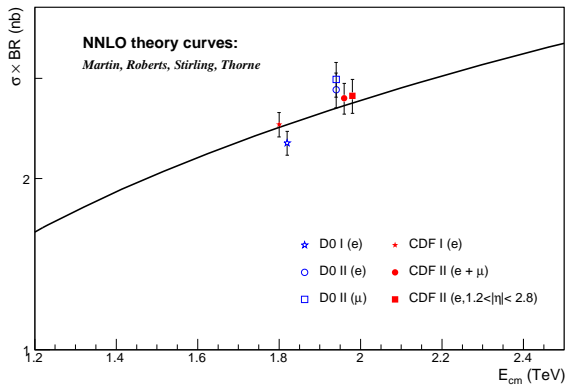}}
\caption{$W$ production cross sections measured by CDF and D0. The
  measurements obtained both at Tevatron I and II agree well with NNLO
  predictions also shown in the plots. The measurements at two different
  center of mass energies have been displaced for clarity.\label{boson}}
\end{figure}

Single production of electroweak bosons represents a benchmark analysis
for D0 and CDF; assuming the validity of SM predictions for its cross
section, it can be used to measure or cross check the luminosity.  This
procedure can be particularly useful at the LHC.  Both for single and
di-boson production\cite{talk_boson} studied in the leptonic decay
channels, the agreement between experimental and theoretical results is
remarkable.  Figure~\ref{boson} shows such good agreement for $W$
production at the Tevatron, both for runs I and II.

A crucial prerequisite for these analyses is the reliability of
predictions from QCD calculations including higher-order corrections.
Calculations at NLO are standard nowadays. Recent studies of $W$ boson
production at large transverse momentum have shown that soft-gluon
corrections at NNLO can be important for a reduction of the residual
scale dependence\cite{talk_kidonakis}. Their inclusion is therefore
essential to obtain predictions with reduced theoretical uncertainties.


\section{Flavor Physics at the B-Factories}

\begin{figure}[t!]
\centerline{\epsfxsize=10cm\epsfbox{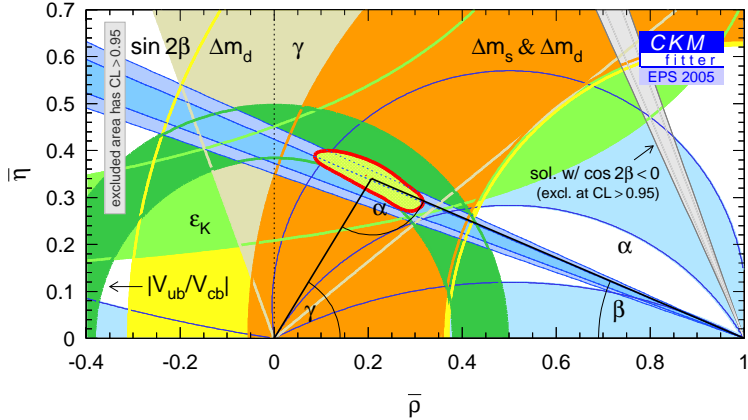}}
\caption{Constraints in the $\rho-\eta$ plane from recent measurements
  at B-factories.\label{ckm}}
\end{figure}

The B-factories at SLAC and KEK are performing very well, continuously
increasing statistics. At the moment, the integrated luminosity is $\sim
560$~fb$^{-1}$ for the Belle collaboration at KEK and $\sim
330$~fb$^{-1}$ for the BaBar collaboration at SLAC. Both experiments
make significant progress in studies of CP violation and rare decays.
The recent precision measurements of the sides\cite{talk_bsides} and the
angles\cite{talk_bangles} of the unitarity triangle have improved the
constraints in the $\rho-\eta$ plane, the two least known parameters of
the Cabibbo-Kobayashi-Maskawa matrix (see Figure~\ref{ckm}). Both Babar
and Belle have also studied a large number of rare
B-decays\cite{talk_bb}, as well as lepton-flavor-violating and rare
decays\cite{talk_btau} of the $\tau$ lepton, observing no deviations
from SM predictions and setting limits on new physics which are
competitive with existing constraints.

\section{Rare Processes}

\subsection{Events with High-$P_T$ Isolated Leptons at HERA}

\begin{table}[b!]
  \tbl{The number of isolated lepton events with $P_T^X > 25$~GeV 
  observed by H1 and ZEUS, compared with the SM predictions.}
  {\footnotesize
\begin{tabular}{ccccc}
                 & Electron & Muon & Combined\\
$P_T^X > 25$~GeV & obs/exp & obs/exp & obs/exp \\[1ex]
\hline
~~H1~~ $e^-p$, 121~pb$^{-1}$ ($98$--$05$) & $2$ / $2.4 \pm 0.5$ & $0$ / $2.0 \pm 0.3$  & $2$ / $4.4 \pm 0.7$ \\
ZEUS $e^-p$, 143~pb$^{-1}$ ($98$--$05$)   & $3$ / $2.9 \pm 0.5$ & ---  & --- \\
\hline
~~H1~~ $e^+p$, 158~pb$^{-1}$ ($94$--$04$) & $9$ / $2.3 \pm 0.4$ & $6$ / $2.3 \pm 0.4$  & $15$ / $4.6 \pm 0.8$ \\
ZEUS $e^+p$, 106~pb$^{-1}$ ($99$--$04$)   & $1$ / $1.5 \pm 0.13$ & ---  & --- \\
\hline
\end{tabular}\label{tab:isol_lep_h1zeus}}
\vspace*{-16pt}
\end{table}

\begin{figure}[ht]
\centerline{\epsfxsize=5.5cm\epsfbox{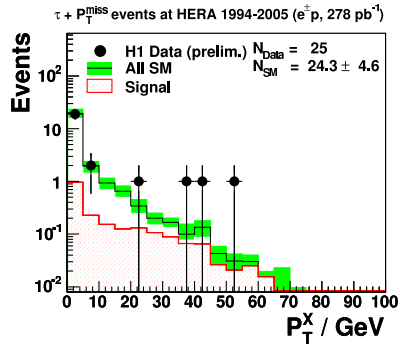}}   
\vspace*{-10pt}
\caption{The hadronic transverse momentum distribution of $\tau +
  P_T^{miss}$ events observed by the H1 experiment is compared to the SM
  expectation. \label{fig:isoltau_h1}}
\end{figure}

In $ep$ collisions at HERA, the production of a $W$ boson has a
cross-section of about $1$~pb. The leptonic decay of the $W$ leads to
events with an isolated lepton (electron, muon or tau) and large missing
transverse momentum. Of particular interest are events where also the
hadronic system has a large transverse momentum ($P_T^X$). The H1
experiment observes an abnormally large rate of such events in the
electron and muon channels\cite{isollep_h1_dis06}. In the analysis of
all HERA-I and HERA-II data, which amount to a total luminosity of
$279$~pb$^{-1}$, $17$ events are observed at $P_T^X > 25$~GeV, whereas
the SM expectation is $9.0 \pm 1.5$. Only $2$ of these events are
observed in $e^-p$ collisions, in agreement with the SM expectation,
while $15$ events are observed in the $e^+p$ data where $4.6 \pm 0.8$
events are expected (see Table~\ref{tab:isol_lep_h1zeus}).
The ZEUS experiment has carried out a re-analysis of the electron
channel using the 1998-2005 $e^-p$ and 1999-2004 $e^+p$
data\cite{isollep_zeus_dis06}, resulting in a larger purity of the
selected $W$ events. The event numbers found in this analysis are also
shown in Table~\ref{tab:isol_lep_h1zeus}. At $P_T^X > 25$~GeV the number
of events observed by ZEUS is in agreement with the SM expectation. A
direct comparison of the results of both experiments remains difficult
since H1 and ZEUS cover different phase-space regions.
A recent analysis of the $\tau$ channel with a total luminosity of
$278$~pb$^{-1}$ was also presented by the H1
collaboration\cite{tau_dis06}. In the $\tau$ channel, the separation of
the $W$ signal from other SM processes is more difficult and the purity
and efficiency are lower than for the $e$ and $\mu$ channels. In total
$25$ events are observed compared to a SM expectation of $24.2 \pm 4.6$.
Three of them have $P_T^X > 25$~GeV where $0.74 \pm 0.19$ events are
expected. The $P_T^X$ distribution is presented in
Figure~\ref{fig:isoltau_h1}.

\subsection{Multi-Lepton Events at HERA}

The main production mechanism for multi-lepton events is photon-photon
collisions. All event topologies with high $P_T$ electrons and muons
have been investigated by the H1 experiment\cite{mlep_dis06} using a
total luminosity of $275$~pb$^{-1}$.  The measured yields of di-lepton
and tri-lepton events are in good agreement with the SM, except in the
tail of the distribution of the scalar sum of $P_T$ of the leptons
($\sum P_T$). In $e^+p$ collisions, $4$ events are observed with $\sum
P_T > 100$~GeV compared to a SM prediction of $0.6 \pm 0.1$. No such
event is observed in $e^-p$ collisions with a similar SM expectation of
$ 0.5 \pm 0.1$.
A first measurement of the cross-section of $\tau$-pair production in
$ep$ collisions was also presented by H1\cite{tau_dis06}.


\section{Searches for New Resonances}

\subsection{New Gauge Bosons}

New resonances coupling to leptons are predicted in several extensions
of the SM. Extending the SM gauge group often leads to additional $Z'$
bosons, with most popular models being based on left-right symmetry or
on the $E_6$ Grand Unification group.
Both D0 and CDF experiments have looked for di-lepton resonances in the
Run II data, including all three channels with di-electrons, di-muons
and di-taus. Good agreement with the SM predictions is observed in the
di-lepton invariant mass distributions, up to the highest
masses\cite{tevatron_searches_dis06}. The highest direct bound on $M_{Z'
  \rightarrow ee} > 850$~GeV is set by CDF using a total luminosity of
$820$~pb$^{-1}$.
The possible production of a $W'$ boson decaying into $e$-$\nu_e$ was
also investigated by CDF\cite{tevatron_searches_dis06}. No signal is
observed and a lower bound on the $W'$ mass of $788$~GeV was obtained.


\subsection{Leptoquarks}

An intriguing characteristic of the SM is the observed symmetry between
the lepton and the quark sectors, which is manifest in the
representation of the fermion fields under the SM gauge group, and in
their replication over the three family generations.  Models which
implement a new symmetry between the lepton and quark sectors lead to
new ``lepto-quark'' interactions.  Leptoquarks (LQs) are new scalar or
vector color-triplet bosons, carrying a fractional electromagnetic
charge and both a baryon and a lepton number. Several types of LQs might
exist, differing in their quantum numbers. The
Buchm\"uller-R\"uckl-Wyler (BRW) classification of LQs is based on the
assumption that LQs have purely chiral couplings to SM fermions and obey
the SM $SU(2)$ symmetry.  The interaction of a LQ with a lepton-quark
pair is of Yukawa or vector type and is parametrised by a coupling
constant $\lambda$.  Depending on its quantum numbers, a LQ couples to
$eq$, $\nu q$ or both.  The branching ratios $\beta$ for a LQ to decay
into $eq$ or $\nu q$ are fixed in the minimal BRW model, but can be
treated as a free parameter, e.g.\ if the LQs mix with other new heavy
particles.

At the Tevatron, leptoquarks are mainly pair-produced via their coupling
to gluons.  First generation LQs have been looked for in Run II data in
the $eejj$, $e\nu jj$ and $\nu\nu jj$ final states ($j$ denotes a jet).
Masses below $256$~GeV for LQs decaying solely into an electron and a
quark are ruled out by the D0 experiment, combining Run I and Run II
data.  In comparison, in deep inelastic scattering, LQs are produced
through their Yukawa-type coupling $\lambda$. HERA
experiments\cite{lqh1_dis06} exclude LQ masses below $\sim 290$~GeV, if
the coupling $\lambda$ is of electromagnetic strength ($\lambda^2/4\pi =
\alpha_{em}$).  The searches at the Tevatron for scalar LQs with $\beta
= 0$ in the $\nu\nu jj$ channels have been updated using events with two
acoplanar jets and missing transverse energy and new mass bounds of
$136$ and $117$~GeV are set by D0 and CDF, respectively.

Leptoquarks coupling to second or third generation fermions and with
masses above $100$~GeV can be directly searched for only at the
Tevatron. Pair-produced LQs leading to final states with two muons or
one muon and missing transverse energy and two jets ($\mu \mu jj$ or
$\mu \nu jj$) have been searched for in the Run II data. For $\beta = 1$
and second generation LQs, mass bounds of $224$~GeV using Run II data
only and $251$~GeV by combining Run I and II data are obtained by the
CDF and D0 experiments.  CDF also looked for third generation LQs in the
$\tau \tau jj$ channel\cite{tevatron_searches_dis06} and found lower
limits of $344$~GeV and $151$~GeV on the masses of vector or scalar LQs,
respectively.


\subsection{Excited Fermions}

\begin{figure}[ht]
\centerline{
\epsfxsize=5.9cm
\epsfbox{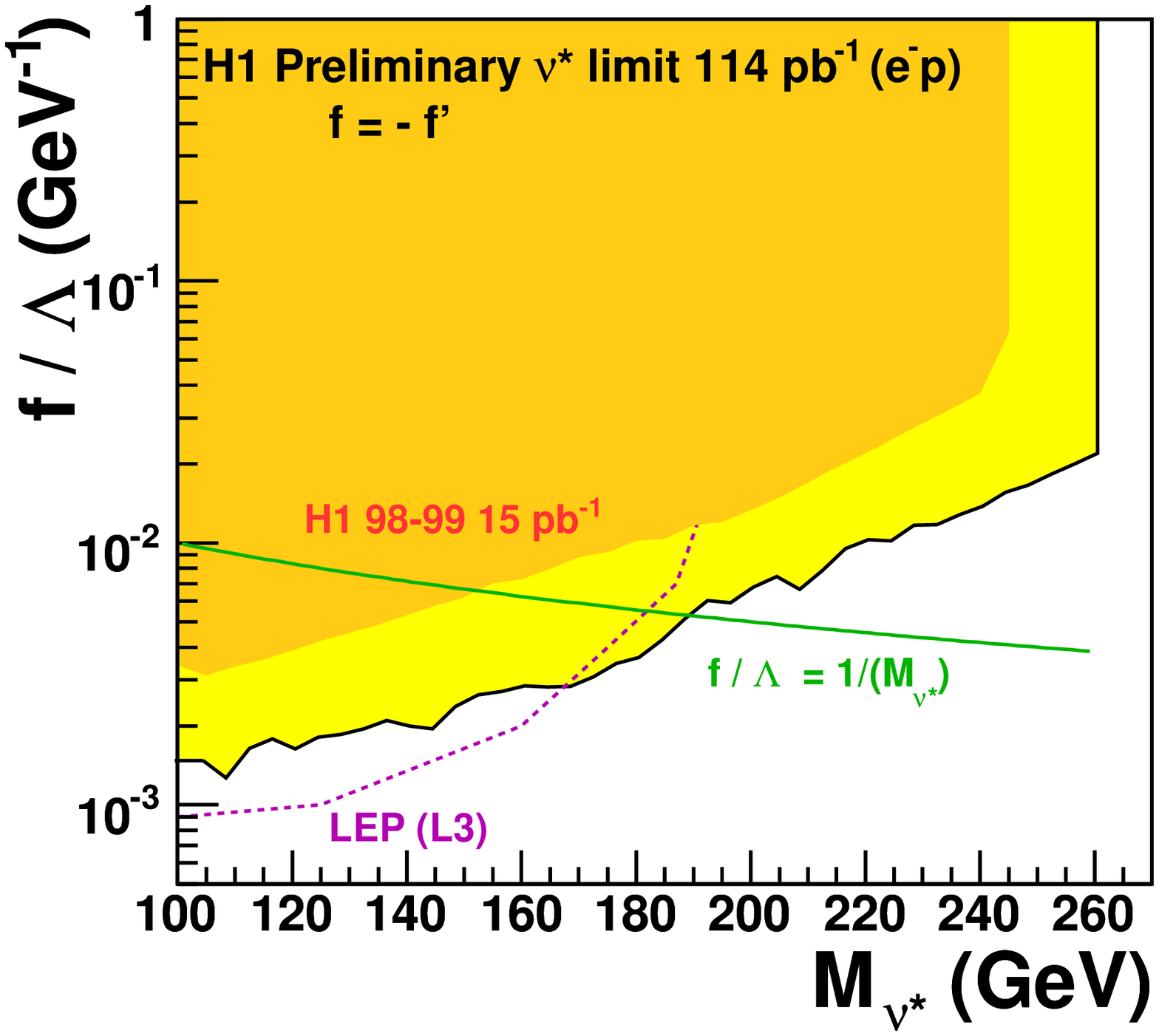}
\put(-80,147) {(a)}
\epsfxsize=5.4cm
\epsfbox{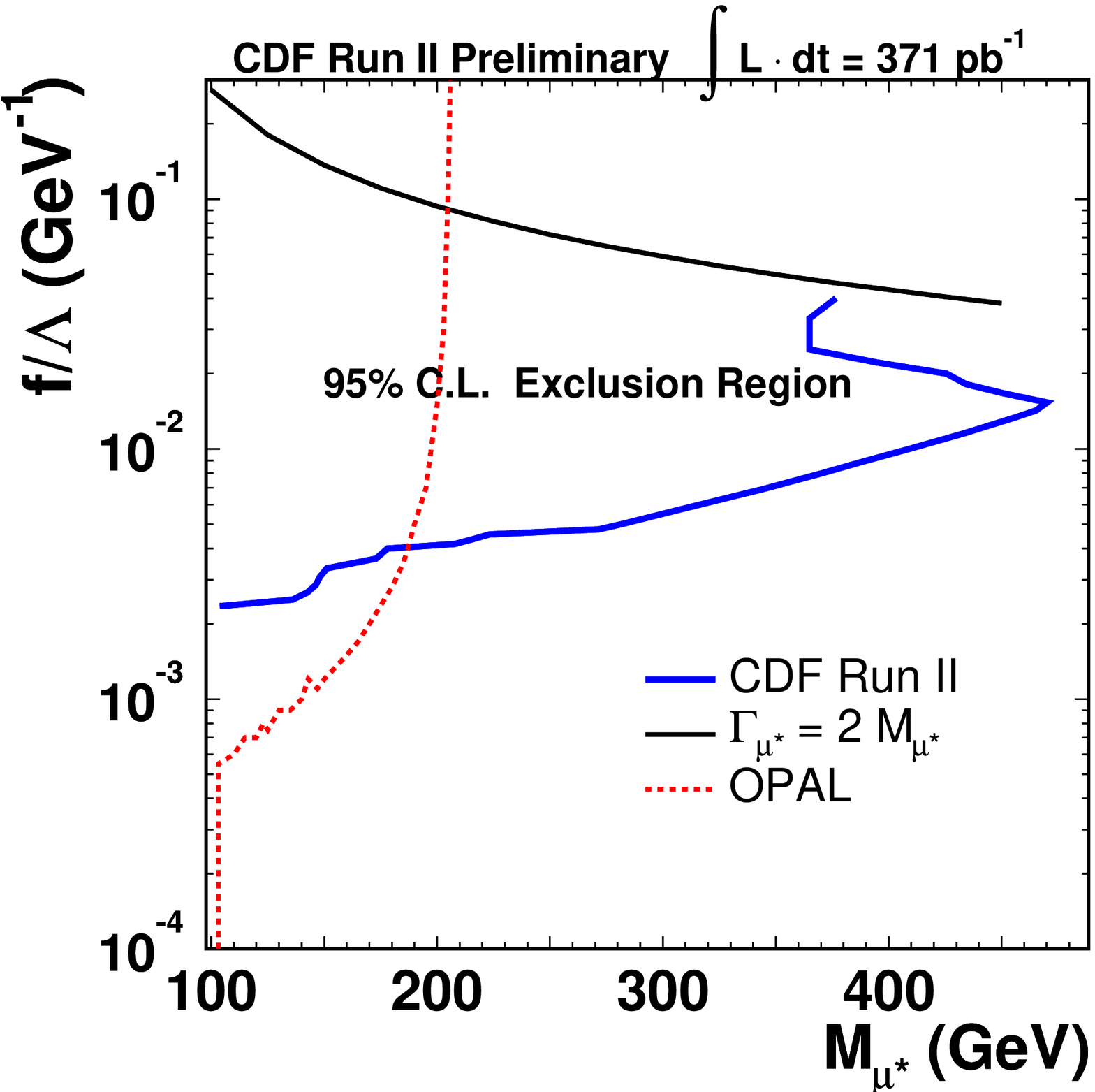}
\put(-75,147) {(b)}}   
\vspace*{-5pt}
\caption{Exclusion limits on the production of excited neutrinos (a)
  and muons (b) obtained by the H1 and CDF experiments,
  respectively.\label{fig:excited_fermion}}
\end{figure}

The observed replication of three fermion families motivates the
possibility of a new scale of matter yet unobserved.  An unambiguous
signature for a new scale of matter would be the direct observation of
excited states of fermions ($f^*$), via their decay into a gauge boson.
The interaction of an $f^*$ with a gauge boson can be effectively
described by a magnetic coupling (i.e., a dimension-five operator)
proportional to $1/\Lambda$ where $\Lambda$ is a new scale.
Proportionality constants $f$, $f'$ and $f_s$ result in different
couplings to $U(1)$, $SU(2)$ and $SU(3)$ gauge bosons.
Recently, the H1 experiment has carried out a search for excited
neutrinos taking advantage of the new sensitivity brought by
$114$~pb$^{-1}$ of HERA-II data from $e^-p$
collisions\cite{nustar_h1_dis06}.  The bound on the $\nu^*$ mass
obtained as a function of $f/\Lambda$ is presented in
Figure~\ref{fig:excited_fermion}a. Assuming $f/\Lambda = 1/M_{\nu^*}$
and $f = - f'$, masses below $188$~GeV are ruled out.
The single production of excited muons $\mu^*$ has been looked for at
the Tevatron by both CDF and D0 experiments. In an analysis based on a
contact interaction formalism for the production and also possibly for
the decay of the $\mu^*$, lower mass limits of $800$~GeV and $618$~GeV
for $\Lambda = 1$~TeV can been set by CDF and D0, respectively.
Figure~\ref{fig:excited_fermion}b presents the corresponding exclusion
region for $M_{\mu^*}$ as a function of $f/\Lambda$ obtained in the CDF
analysis if the same model as the one used by H1 is considered.
The D0 experiment also looked for heavy resonances in the $Z$+jet
channel as a signal of excited quark production. Assuming a production
and decay mechanism via contact interactions, $q^*$ masses below
$520$~GeV have been excluded\cite{tevatron_searches_dis06}.

\section{Supersymmetry}

\begin{figure}[t!]
\centerline{\epsfxsize=5.2cm
\epsfbox{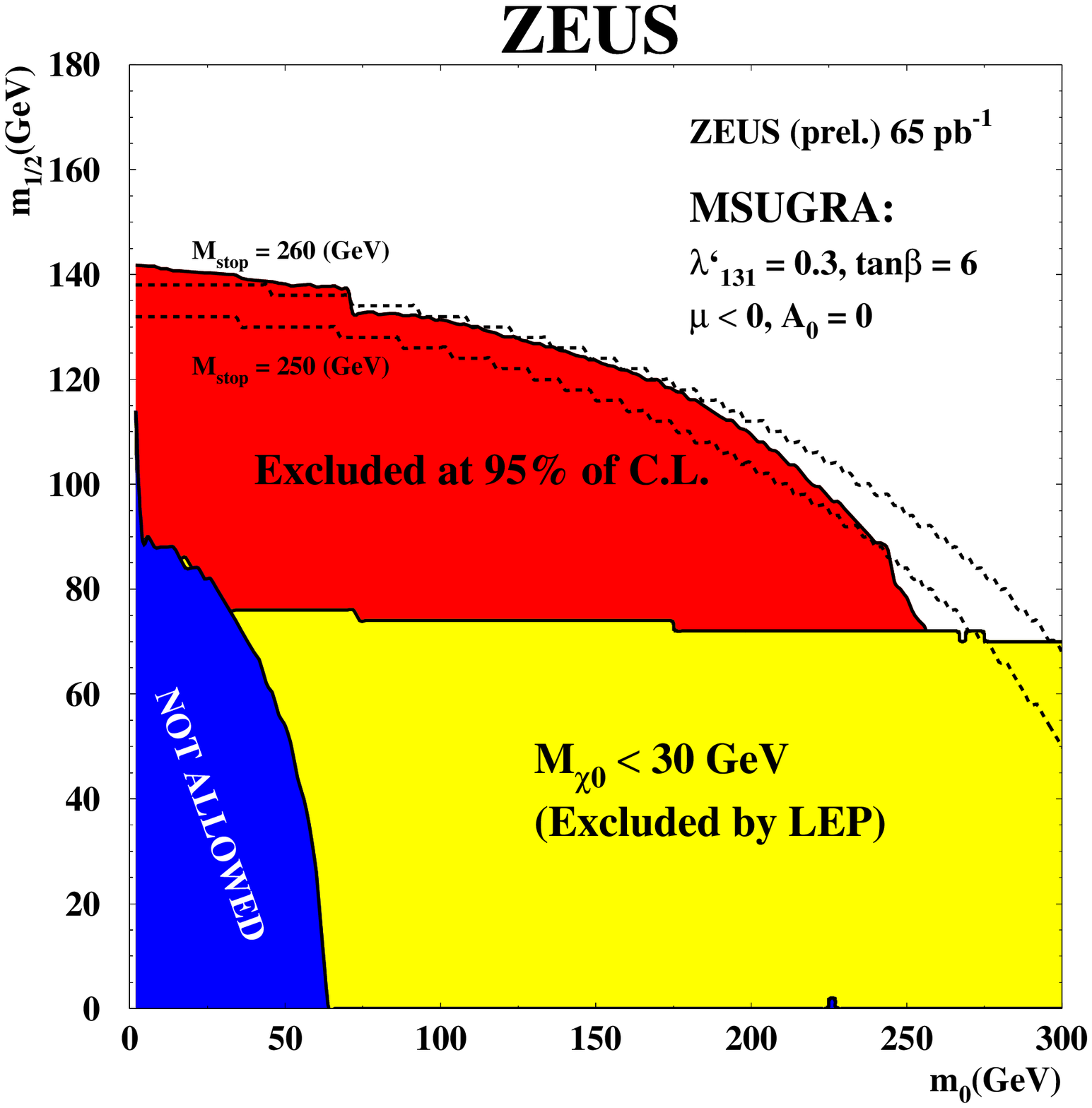}
\put(-27,138) {(a)}
\epsfxsize=5.1cm
\epsfbox{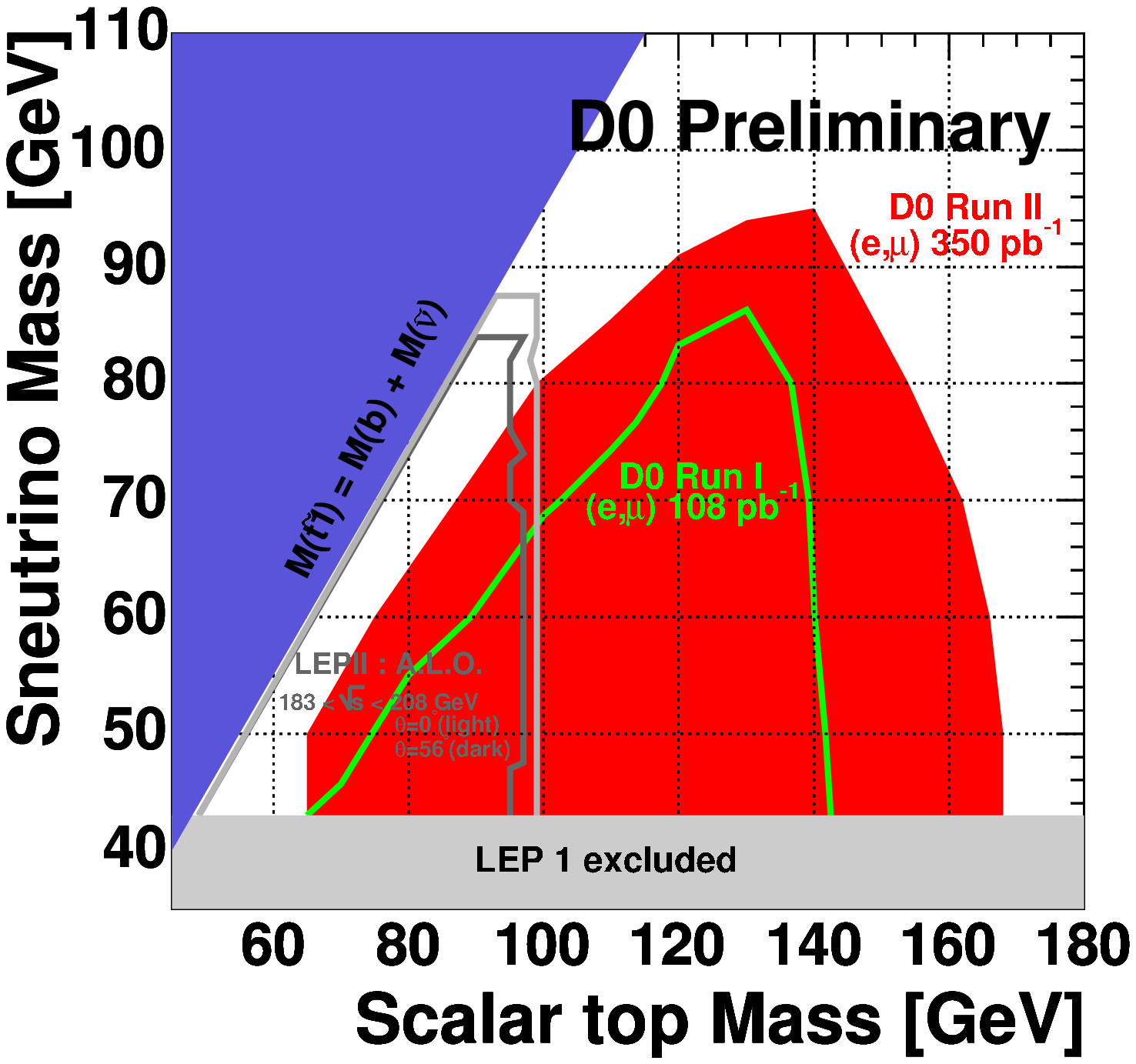}
\put(-20,138) {(b)}}   
\vspace*{-5pt}
\caption{Stop quark production: parameter space regions excluded by
  ZEUS (a) and D0 (b) in $R_p$-violating and $R_p$-conserving scenarios,
  respectively. \label{fig:stop}}
\end{figure}

A popular extension to the SM is supersymmetry (SUSY). SUSY unifies
internal symmetries with Lorentz invariance and associates
supersymmetric partners ($s$particles) to the known SM particles.
Supersymmetric models provide solutions to many problems of the SM
(hierarchy, fine--tuning, unification) and predict spectacular final
states in high-energy particle collisions. Despite extensive studies at
colliders and elsewhere, no unambiguous signal of SUSY has yet been
detected.
The production of single $s$particles is possible if the conservation of
the multiplicative quantum number $R_p$ is violated ($R$-parity is $R_p
= (-1)^{3B+L+2S}$ where $B$, $L$ and $S$ denote the particle's baryon
number, lepton number, and spin, respectively). In $R$-parity violating
models, $s$-channel $s$quark production at HERA via the
electron-quark-$s$quark Yukawa coupling ($\lambda$) is possible. A
special case is the $s$top ($\tilde{t}$) which in many SUSY scenarios is
the lightest $s$quark.
A recent search for $\tilde{t}$ production at HERA was reported by the
ZEUS experiment\cite{zeus_susy_dis06}. The analysis involved both
$R_p$-violating and gauge decays of the $s$top and is therefore
sensitive to a large class of models, corresponding to a wide scan over
the parameter space. The exclusion limits obtained are in agreement with
previous results from H1 and scenarios with $s$top quarks with mass
below $\sim 260$~GeV can be excluded if $\lambda ' = 0.3$ (see
Figure~\ref{fig:stop}a).

Searches where $R_p$ is conserved are mainly based on the presence of
the lightest supersymmetric particle in the final state, which usually
involves large missing transverse momentum. In this case, SUSY particles
can only be produced in pairs.
A search for the production of $s$top pairs has been performed by D0,
considering scenarios where the lightest $s$quark is a $s$top
$\tilde{t}_1$. Final states with $e^\pm \mu^\mp + b \bar{b} +
E_T\!\!\!\!\!\!\!\!\diagup\;$ and $\mu^\pm \mu^\mp + b \bar{b} +
E_T\!\!\!\!\!\!\!\!\diagup\;$ have been
investigated\cite{tevatron_susy_dis06}, but no deviations from the SM
expectations were observed. The excluded region is presented in
Figure~\ref{fig:stop}b.
Looking for the possible production of charginos and neutralinos in
mSUGRA scenarios where charginos and neutralinos decay leptonically, a
large variety of final states with three leptons and missing energy have
been investigated by CDF\cite{tevatron_susy_dis06}. Integrated
luminosities of $\sim 700$~pb$^{-1}$ are used but no significant excess
is observed.

If Supersymmetry is realized in Nature, we may expect that experiments
at the Tevatron and the LHC will announce its discovery in the near
future. For high-precision analyses of supersymmetric parameters with
the aim to reconstruct the fundamental supersymmetric theory, one will
have to wait for the completion of high-luminosity runs at the LHC and
the contribution from experiments at the planned ILC\cite{talk_SPA}.
Whereas algorithms for SUSY analyses of future experimental results have
been worked in great detail, the discovery of non-SUSY new physics will
require the development of new strategies\cite{talk_knuteson}.


\section{Conclusions}

Despite the impressive amount of precision electroweak measurements, no
clear sign of deviations from SM expectations has been observed.
Nevertheless important discoveries seem to be quite close.  The
experimental results strongly indicate a light Higgs boson; its possible
mass range will be accurately scanned in the next few years both by the
Tevatron and LHC experiments hopefully clarifying the status of the SM
Higgs sector.  The precision measurements of the B-factories, which will
double their statistics in the next couple of years, will also
contribute in testing the SM by means of CP-violation and rare decays
studies.



\end{document}